\documentclass[prl,showpacs,twocolumn,floatfix]{revtex4}

\usepackage{graphicx} 
\usepackage{dcolumn}  
\usepackage{bm}       
\usepackage{amssymb}  
\usepackage{amsmath}  

\input epsf
\input rotate

\def\al{_\alpha}
\def\fal{_{f,\alpha}}

\def\bei{\begin{itemize}}
\def\eei{\end{itemize}}
\def\alNf{_{\alpha=1}^{N_f}}

\def\f{_f}
\def\0{^{(0)}}

\def\bea{\begin{eqnarray}}
\def\eea{\end{eqnarray}}
\def\ben{\begin{equation}}
\def\een{\end{equation}}
\def\benu{\begin{enumerate}}
\def\enu{\end{enumerate}}

\def\n{n}
\def\lsim {\ifmmode {\buildrel<\over\sim}}

\def\sss{\scriptscriptstyle\rm}

\def\1var{(\bx_1...\bx\N)}

\def\br{{\bf r}}

\def\b1{{\bf 1}}
\def\bx{{x}}

\def\s{_{\sss S}}

\def\Hxc{_{\sss HXC}}

\def\N{_{\sss N}}

\def\ee{_{\rm ee}}

\def\sph_int{ {\int d^3 r}}

\def\JCP{J. Chem. Phys.\ }

\def\infintd3r{ \int_{-\infty}^\infty d^3r\,}
\def\intd3r{ \int d^3r\,}

\def\laplace1d{\frac{d^2}{dx^2}}
\def\plaplace1d{\frac{d^2}{d{x'}^2}}

\def\padr2{\frac{\partial^2}{\partial r^2}}

\def\N{{\cal N}}

\def\b{{\beta}}

\begin{document}

\headheight 50pt

\title{Partition Density Functional Theory}
\author{Peter Elliott and Kieron Burke}
\affiliation{Departments of Physics and of Chemistry, University of California, Irvine, CA 92697}
\author{Morrel H. Cohen}
\affiliation{Department of Physics and Astronomy, Rutgers University, 136 Frelinghuysen Road, Piscataway, NJ 08854 \\ Department of Chemistry, Princeton University, Washington Road, Princeton, NJ 08544}
\author{Adam Wasserman}
\affiliation{Department of Chemistry, Purdue University, West Lafayette, IN 47907}


\begin{abstract}
Partition density functional theory is a formally exact
procedure for calculating molecular properties from 
Kohn-Sham calculations on isolated fragments, interacting via
a global partition potential that is a functional of the fragment densities.
An example is given and consequences discussed.
\end{abstract}


\date{\today}

\maketitle

Kohn-Sham density functional theory (KS-DFT)\cite{HK64,KS65} 
is an efficient and usefully accurate electronic structure method,
because it replaces the interacting
Schr\"odinger equation with a set of single-particle orbital
equations. Calculations with several hundred atoms are now routine, but
there is always interest in much larger systems.
Many such systems are treated by a lower-level
method, such as molecular mechanics,
but a fragment in which a chemical reaction
occurs must still be treated quantum mechanically.  A plethora of such
QM/MM approaches have been tried and tested, with varying degrees of
success\cite{plethora}.  These are often combined with attempts
at orbital-free DFT, which avoids the KS equations,
but at the cost of higher error and unreliability.

On the other hand, partition theory (PT) \cite{CW07,CW06}
combines the simplicity of functional minimization with a density
optimization to define fragments (such as atoms) within molecules,
overcoming limitations of earlier approaches to reactivity
theory\cite{PDLP78,AP01}. 
While there are now many definitions of, e.g., charges on atoms,
none have the generality of PT and the associated promise of
unifying disparate chemical concepts. However, previous work
on PT has been either formal\cite{CW07,CW06} or for two atom systems\cite{CWB07,CWCB09}.

In this paper, we unite KS-DFT with PT to produce an
algorithm that allows a KS calculation for a molecule to
be performed via a self-consistent loop over isolated fragments.
Such a fragment calculation \emph{exactly} reproduces the result of
a standard KS calculation of the entire molecule.
We demonstrate its convergence on a 12-atom example.
This also shows that fragments can be calculated 'on the fly', as part of
solving any KS molecular problem.

Thus we present a formally exact framework within which existing
practical approximations can be analyzed and, for smaller systems, 
compared with exact quantities.
In practical terms, our method suggests new approximations
that can, by construction, scale linearly\cite{Y91}
with the number of fragments (so-called $O(N)$), and allow embedding of KS calculations within cruder
force-field calculations (QM/MM). It also suggests ways to improve XC
approximations so as to produce correct dissociation of molecules \cite{CMY08}.


To understand the relation between DFT and PT, recall
that the Hohenberg-Kohn theorem proves that for a given
electron-electron interaction and statistics, the 
external (one-body)  potential $v(\br)$ is a
unique functional of the density $\n(\br)$. 
The total energy can be written as:
\ben
E[\n] = F[\n] + \int d^3r\, \n(\br)\, v(\br),
\een
where $F[n]$ is a universal functional,
defined by the Levy-Lieb constrained search\cite{L79} over
all antisymmetric wavefunctions $\Psi$ yielding density $\n(\br)$:
\ben
F[n] = \min_{\Psi\rightarrow\n(\br)}\langle\Psi |\,  \hat{T} + \hat{V}\ee\, |\Psi\rangle ,
\een
where $\hat{T}$ and $\hat{V}\ee$ are the
kinetic energy and Coulomb repulsion operators respectively.
The KS equations are single-particle equations defined to reproduce $\n(\br)$.
Define the KS energy as
\ben
E\s[\n]=\langle\Phi\s[\n]| \hat{T} + \hat{V} |\Phi\s[\n]\rangle
= T\s[n] + \int d^3r\, \n(\br)\, v(\br) ,
\een
where $\hat{V}$ is the external potential operator, $\Phi\s[\n]$ is the
KS wavefunction (usually a single Slater determinant)
of density $\n(\br)$, and $T\s[\n]$ is the kinetic energy of $\Phi\s[\n]$.
Define the Hartree-exchange-correlation energy, as
\ben
\label{EHxc}
E\Hxc[\n] = E[\n] - E\s[\n],
\een
so that the KS potential is $v(\br) + v\Hxc(\br)$, where
\ben
v\Hxc(\br) = {\delta E\Hxc[n]}/{\delta\n(\br)}.
\een

Partition theory deals with the problem of dividing
a system into localized fragments. 
For molecules or solids, 
\ben
v(\br) = \sum_\beta \ \frac{Z_\beta}{|\br-\mathbf{R}_\beta|}
= \sum\alNf v\al(\br) ,
\een
where $Z_\beta$ is the atomic charge of a nucleus at
point $\mathbf{R}_\beta$, and these are regrouped
into $N\f$ fragment potentials, $v\al(\br)$.
The fragmentation is chosen based on the particular use of PT: e.g., one might
atomize an entire molecule, or merely separate off a well-known chemical species.
The partition problem
is then to divide $\n(\br)$ between the fragments.
There are many methods for doing so, but in PT\cite{CW06,CW07} 
we minimize the total fragment energy
\ben
\label{Ef}
E\f = \min_{\substack{\{\n\al\}\\ \sum\alNf \n\al(\br)=\n(\br)}}\sum\alNf \left(F[\n\al] + \int d^3r\, \n\al(\br)\, v\al(\br) \right) ,
\een
where $\n\al(\br)$ is the density on the $\alpha$-th fragment.
Each fragment is considered to be in contact with a distant reservoir
of electrons\cite{PPLB82}, so its integral $N\al$ need not be an integer.
If $N\al=p\al+\nu\al$, with $0 \leq \nu\al \leq 1$, then\cite{PPLB82}
\ben
\label{epsadf}
F[\n\al] = (1-\nu\al)F[\n_{p\al}] + \nu\al F[\n_{p\al+1}],
\een
where
\ben
\n\al(\br) = (1-\nu\al)\n_{p\al}(\br) + \nu\al\n_{p\al+1}(\br),
\een
and the integer densities are ground-states of a common potential.
Minimizing the Lagrangian:
\ben
{\cal G} = E\f + \int d^3 r \ v_p(\br)\left( n(\br) - \sum\alNf\n\al(\br)\right) ,
\een
yields the correct fragment densities. 
The Lagrange multiplier $v_p(\br)$ is called the partition potential, and satisfies
\ben
\label{frageul}
\frac{\delta F[\n\al]}{\delta\n\al(\br)} +v\al(\br) + v_p(\br) = \mu.
\een
So $\n\al(\br)$ is the ground-state density of $N\al$ electrons in
\emph{effective fragment
potential} $ v\fal(\br) = v\al(\br) + v_p(\br)$. 

Thus PT replaces a molecule of interacting fragments with an effective system
of non-interacting fragments, and $v_p(\br)$ is the analog of $v\Hxc(\br)$
in KS theory.  Analogous to Eq. (\ref{EHxc}), we define the
partition energy as
\ben
\label{Ep}
E_p[\{\n\al\}] = E[\n] - E\f[\{\n\al\}],
\een
which, via Eq. (\ref{Ef}), is a functional of $\{\n\al(\br)\}_{\alpha=1...N}$ . Functional
differentiation yields:
\ben
\label{vp}
v_p(\br) = \delta E_p[\{\n\al\}]/\delta\n\al(\br).
\een
Analogous to KS-DFT, 
once $E_p[\{\n\al\}]$ is given, we have
a closed set of equations that yield the molecular
density and energy at self-consistency. 
Next, we use a superscript
0 to denote quantities evaluated in the physical dissociation limit, where all bond lengths 
between fragments have been
taken to $\infty$ while keeping intrafragment distances fixed.
In this limit, the fragments
do not interact and their densities do not overlap, and $E\f^{(0)}$
is the sum of the truly separated fragments, each with density 
$\n_\alpha^{(0)}(\br)$.
Defining the relaxation energy $E_{rel}=E\f^{(0)}-E\f$, we write
\ben
E_p = E_{dis} + E_{rel},~~~~E_{dis} = E-E\0 ,
\label{E_p}
\een
where $E_{dis}$ is 
the electronic contribution to the dissociation energy.
Thus $E_p < 0$ for any bound molecule (by construction), is expected to be
much smaller than the total energy (on the scale of chemical bonding),
and vanishes as the fragments are pulled apart.

In a KS fragment calculation, the KS potential for the $\alpha$-th fragment is
found from Eq. (\ref{vp}) in KS quantities:
\ben
\label{vsf}
v_{{\sss S},f,\alpha}[\n\al,\bar\n\al](\br) = v\s[\n\al](\br) + \left( v(\br)+
v\Hxc[\n](\br) - v\s[\n](\br) \right),
\een
where $v\s[\n](\br)=- \delta T\s[\n]/\delta\n(\br)$, and $\n(\br)=\n\al(\br)+\bar\n\al(\br)$.
This is the central result of this paper, as it gives the
fragment KS potential for a pair of trial densities, $n\al(\br)$ and $\bar\n\al(\br)$,
in terms of quantities from KS-DFT.

\begin{figure}
\begin{center}
\unitlength1cm
\begin{picture}(4,4.8)
\put(-4.1,-3.2){\makebox(4,4.8){
\includegraphics{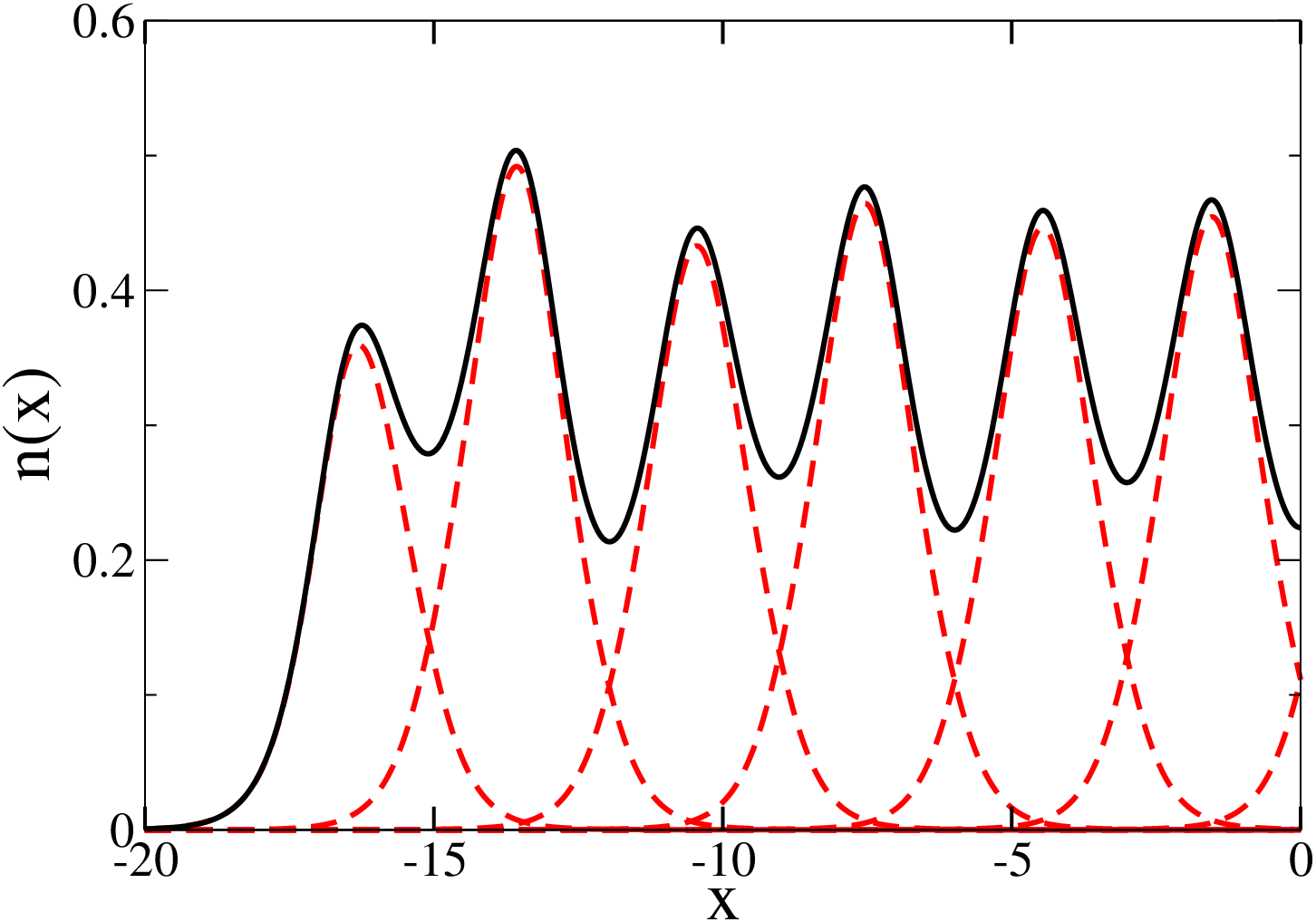}
}}
\end{picture}
\end{center}
\caption{Solid line: The exact spin-unpolarized ground state of 12 electrons in the potential of Eq. (\ref{pot}). Dashed lines: The fractionally occupied fragment densities. By symmetry, the other half of the density is simply the mirror image of that shown.}
\label{f:frag_den}
\end{figure}

In Eq. (\ref{vsf}), $v\s[\n\al](\br)$ is simply the KS fragment potential from the
previous iteration, but $v\s[\n](\br)$ is the KS potential for a trial density for
the whole molecule.  Many methods exist for finding this\cite{inversion}.
We iterate\cite{PVW03}:
\ben
\label{invdfpt}
v\s^{(m+1)}(\br) = v\s^{(m)}(\br) + \gamma\left[ n^{(m)}(\br) - n^{(k)}(\br) \right] ,
\een 
where $n^{(m)}(\br)$ is the density found from potential $v\s^{(m)}(\br)$, $\gamma > 0$ is a constant, and
$n^{(k)}(\br)$ is the target density (sum of fragment densities from the $k$'th PDFT
iteration) whose KS potential we are trying to find.
To find the fragment occupations, note that at self-consistency, 
the chemical potentials of all the fragments will be equal. 
We choose $ N\al^{(k+1)} = N\al^{(k)} -\Gamma\left( \mu\al^{(k)} - \bar{\mu}^{(k)} \right)$,
where $\Gamma$ is another positive constant and $\bar{\mu}$ is the average of the fragment chemical potentials, 
used in conjunction with Eq. (\ref{epsadf}) for the functionals\cite{foot}.

The starting point in PDFT is to solve the KS equations for
each isolated fragment, generating their self-consistent
densities $\n^{(0)}\al(\br)$ and KS potentials, and a trial molecular density which
is the sum of overlapping atomic densities.   Then Eq. (\ref{invdfpt})
is iterated to find its KS potential, which completes the inputs for Eq. (\ref{vsf}),
and the cycle repeated to self-consistency, when $E_{dis}$
is found by evaluating $E_p$
on the final fragments, and subtracting the relaxation energies via Eq. (\ref{E_p}).

\begin{figure}
\begin{center}
\unitlength1cm
\begin{picture}(4,4.8)
\put(-4.1,-3.2){\makebox(4,4.8){
\includegraphics{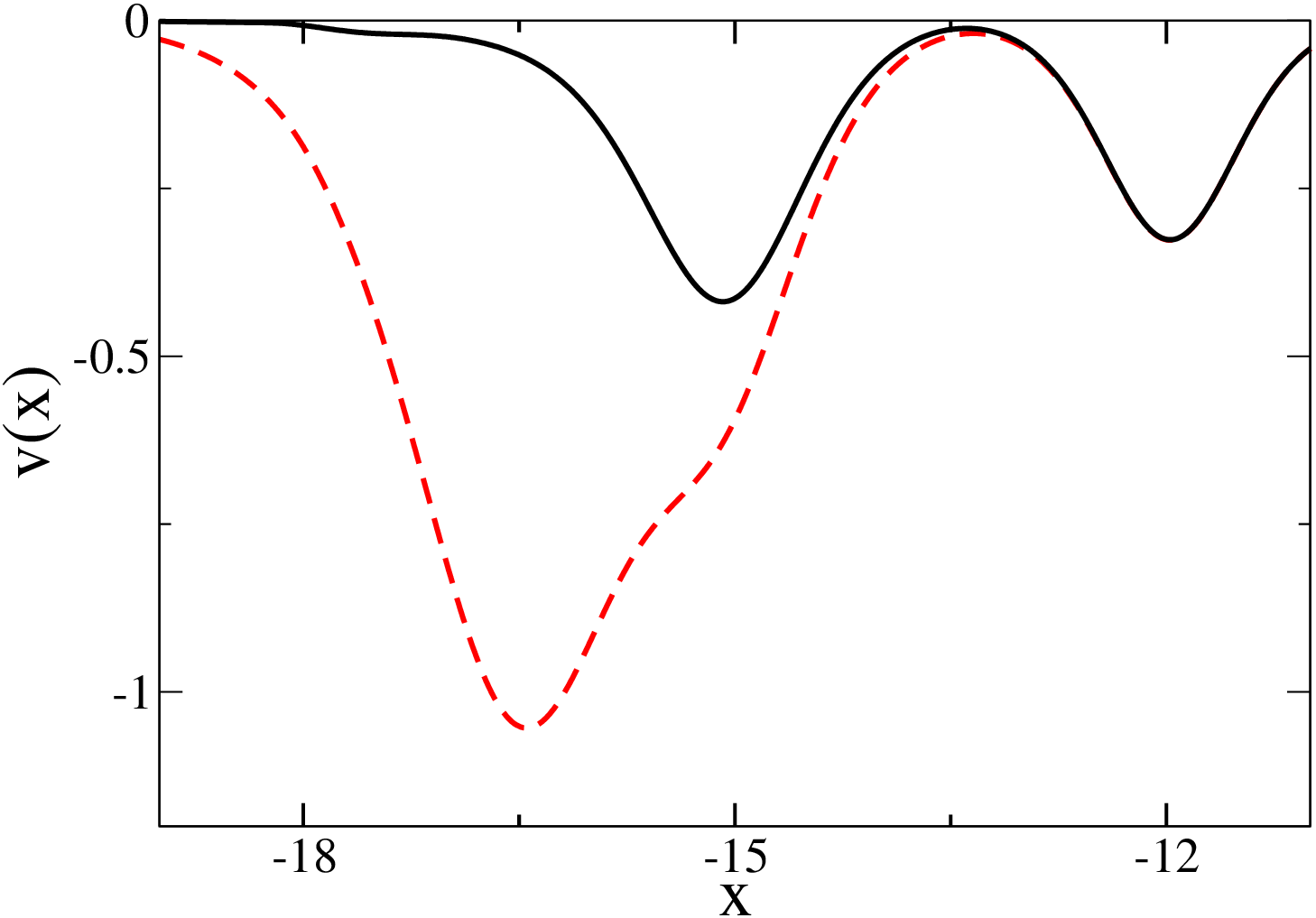}
}}
\end{picture}
\end{center}
\caption{The exact partition potential (solid line) for the atomized chain and the fragment potential for the last atom (dashed line). The ground state with an occupation of $0.77$ in this potential can be seen as the end fragment density in Fig 1.}
\label{f:zoom_fragden}
\end{figure}

\begin{figure}
\begin{center}\unitlength1cm
\begin{picture}(4,4.8)
\put(-4.1,-3.2){\makebox(4,4.8){
\includegraphics{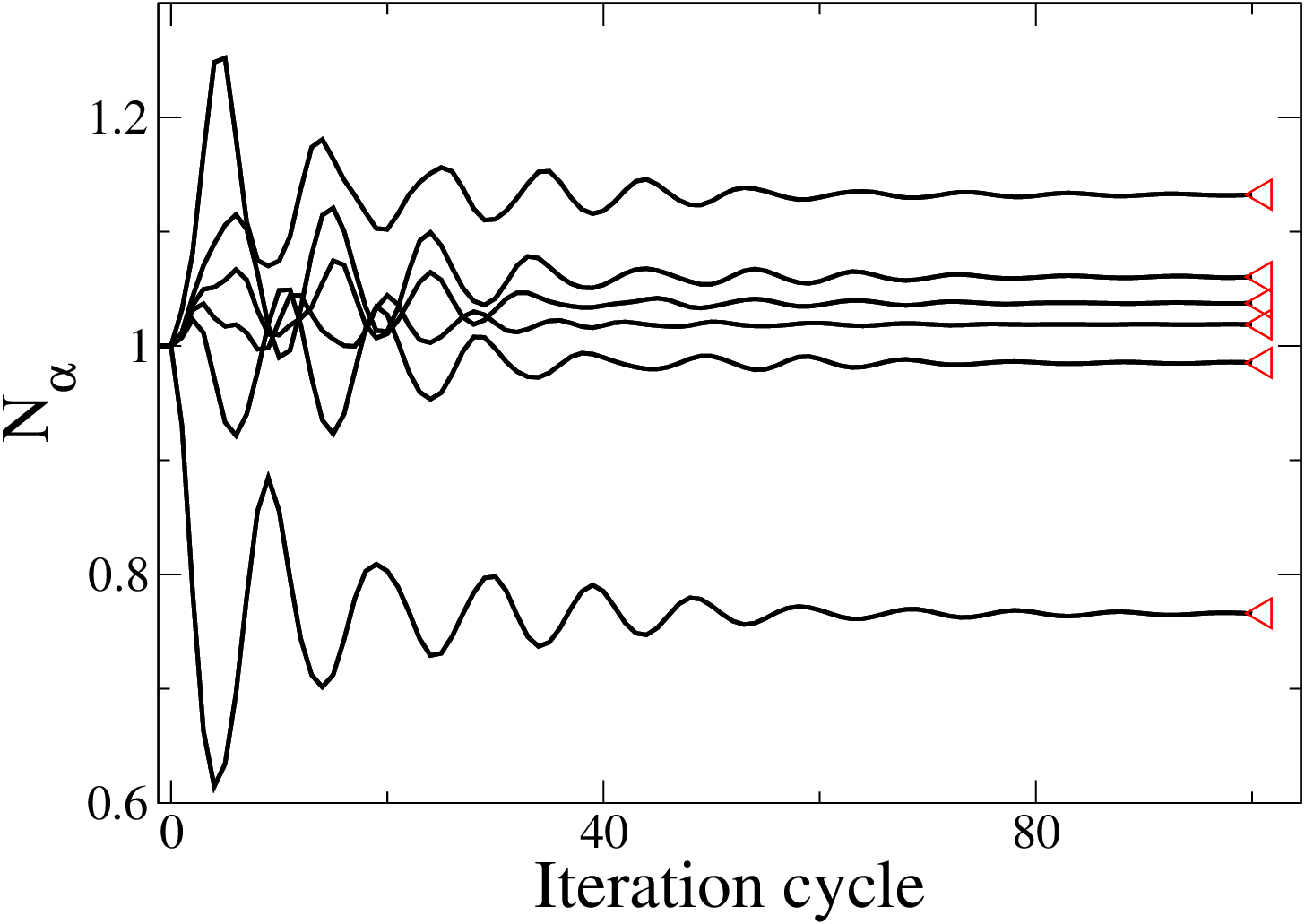}
}}
\end{picture}
\end{center}
\caption{The convergence of the fragment occupation values, $N\al$, during an exact PDFT calculation.
}
\label{f:n12_cvg}
\end{figure}

To show that our algorithm converges, we performed
a PDFT calculation on a $12$-atom $1$d chain with $12$
spin-unpolarized non-interacting fermions, with potential:
\ben
\label{pot}
v(x) = \sum_{\alpha=1}^{12} \frac{-1}{\cosh^2[x+(\alpha - 6.5)R]}.
\een
We chose complete atomization into 12 fragments, so we only ever solve
one- or two-electron problems in a single well.
Fig. \ref{f:frag_den} shows the atomic and molecular 
densities after convergence. The molecular density is identical to that found by direct solution
of the eigenvalue problem for the entire molecule, and doubly occupying
the first 6 eigenstates, which are delocalized over the entire molecule.
We see a small alternation between higher and lower densities throughout
the molecule.  The fragment density occupations reflect this, being
0.77,1.13,0.98,1.06,1.02,1.04 moving inwards towards the center of the chain. 
In Fig. \ref{f:zoom_fragden}, we show both the 
partition potential and effective fragment
potential for the last atom. The (not very large) $v_p(\br)$ polarizes 
the density toward the molecular center, and shifts
the density inwards compared to a free atom.
The partition potential continues throughout the whole chain,
lowering each fragment potential in the bonding region between atoms.
The depth of these troughs oscillates, reflecting the oscillation in occupations.
In Fig. \ref{f:n12_cvg}, we show the convergence of the occupation numbers
to their final values, after some initial oscillations. The total energy of the molecular system can be found via Eq. (\ref{Ep}). We find $E_f = -5.888$ and $E_p = -1.803$ leading to $E = -7.691$, which is exactly that of the direct solution. Since $E^{(0)}_f$ is $-6$, $|E_{rel}| \ll |E_p| \ll |E|$, as expected.

Our calculation was in fact far
more expensive than a regular KS calculation, because we invert the
KS problem for each trial molecular density {\em exactly}.
But the purpose here was not speed, but 
the calculation of exact partition potentials
for small molecules and simple solids.  It produces the {\em exact} partition
potential corresponding to a given KS calculation for the molecule.

The many potential uses of PDFT are made clear by this example.
In principle, Eq. (\ref{vsf}) is exact, but requires the KS potential of
the entire system and to deduce the energy at the end of the
calculation, one needs
\ben
E_p = \Delta T\s[{\n_\alpha}]+ \Delta E\Hxc[{n_\alpha}] + \sum_{\alpha,\beta\neq\alpha}^{N_f}\int d^3r\,\n\al(\br) v_\beta(\br) ,
\een
where $\Delta G[{\n_\alpha}] = G[\sum \n_\alpha]-\sum G[\n_\alpha]$.
However, any local-type approximation makes the method
$O(N)$.  Thus, all the attempts of orbital-free DFT, to find useful approximations
to $\Delta T\s[\n]$, have now a simple framework in which to be tested\cite{BDKW08}.  Moreover, 
there are no formal difficulties arising from
taking density variations within a fixed density, as the trial molecular density
is simply the sum of the fragment densities, which are varied freely.
Although the exact fragment $T\s$ and $v\s(\br)$ would be known during a
calculation, approximations for $\Delta T\s$ would take full advantage
of any cancellation of errors.
For embedding calculations,  a simple approximation would be to treat the
system plus some fraction of its environment (a border region) exactly, and
all the rest approximately.  Since the KS potential is typically near-sighted,
such a scheme should converge rapidly.

For the dissociation of molecules, one can also see how to ensure correct
dissociation energies within PDFT:  simply constrain occupations to be
those of the  isolated fragments.  For H$_2$, we constrain the spin
occupations on the fragments to be (0,1) and vice versa.  Of course,
this is what happens when symmetry is broken as the bond is stretched,
and the difficulty is in producing a scheme that seamlessly goes over
to (1/2,1/2) occupations as $R$ reduces to the equilibrium value.
The value of our formalism is that it produces a framework
for both addressing these questions and constructing approximate solutions.

There is a simple adiabatic connection formula for PDFT.   Consider
scaling all bond lengths between fragments by $\lambda^{-1}$ (again
keeping intrafragment densities fixed), where
$0 < \lambda \leq 1$.  For each $\lambda$, we find those molecular
densities whose fragment densities match those of our molecule, and
define the corresponding partition energy, $E_p(\lambda)$. 
At $\lambda=1$, we have the original molecule; as $\lambda\to 0$,
the bonds become large and the fragments do not interact, so that $E_p(0)=0$.
For intermediate $\lambda$, the molecular density is simply that of the
fragments, overlapped a distance $R/\lambda$ apart.
Then
\ben
E = E_f + \int_0^1 d\lambda\, \frac{dE_p(\lambda)}{d\lambda}.
\een
This allows all the methods of traditional intermolecular 
symmetry-adapted perturbation theory (SAPT)\cite{MJS03} to
be applied to this problem, but with the advantage that the
fragment densities remain fixed.  Interestingly, because the
fragments will generally have dipole moments, the partition
energy decays as $1/R^3$, so that the integrand above behaves
as $\lambda^2$.  (For physical systems that are well-separated
and have attractive van der Waals forces, such effects must
be cancelled by analogous terms in $E_{rel}$). 

There has been considerable previous work on schemes designed
to allow a fragment calculation of a larger molecule, either
within the framework of orbital-free DFT or atomic deformation
potentials, sometimes producing the same (or similar) equations.
Among the earliest, Cortona's crystal potential 
(later called embedding potential)\cite{C91, C92} is an intuitive
prescription for $v_p(\br)$.  
But our formalism reproduces the {\em exact}
solution of the original problem, using only quantities that are
already defined in KS-DFT.  For example, this is not possible
in general without the ensemble definition
of Eq. (\ref{epsadf}), which produces the correct self-consistent
occupations (unlike, e.g., the self-consistent atomic deformation method\cite{scad,MSB96},
where this choice leads to a basis set dependence\cite{OBMP03}) .
We also never freeze the total density\cite{WW93,GWC99,HC06}, but only ever consider it
as a sum of fragment densities.  This avoids ever needing
density variations that are limited by some frozen total density,
which produces bizarre functional derivatives, different from those of KS DFT.
None of these issues arise once smooth (e.g. local or
gradient-corrected) approximations are made to the kinetic
energy functional\cite{C91,C92,MSB96}, but they are vital in 
a formally exact theory.
Thus the present PDFT can be regarded as a
formal exactification (and therefore justification)
of these pioneering
works.

KB and PE acknowledge support under NSF grant CHE-0809859.
KB thanks Filipp Furche for useful discussions.


\begin{references}
\bibitem{foot}
Our algorithm also applies within pure PT, to find $v_p(\br)$ for a known $\n(\br)$.
Replace $v\s(\br)$ by $v_p(\br)$ in Eq. (\ref{invdfpt}), $n^{(k)}(\br)$ is the fixed molecular density, and
$n_f^{(n)}(\br)$ is the sum of fragment densities found from
individual calculations on the 
fragments using $v_p^{(n)}(\br)$ to construct the fragment potentials.
\end{references}
\end{document}